\PassOptionsToPackage{prologue,table}{xcolor}
\documentclass[acmtog]{acmart}
\acmSubmissionID{2088}

\usepackage{booktabs} %

\definecolor{best1}{RGB}{222,242,212}
\definecolor{best2}{RGB}{255,250,212}
\definecolor{best3}{RGB}{255,210,210}
\newcommand{\abbtitle}{TaoGS\xspace}
\usepackage{soul}
\definecolor{junglegreen}{rgb}{0.113, 0.639, 0.5}

\usepackage[ruled]{algorithm2e} %

\SetAlFnt{\small}
\SetAlCapFnt{\small}
\SetAlCapNameFnt{\small}
\SetAlCapHSkip{0pt}

\usepackage{multirow}
\usepackage{colortbl}
\usepackage[table]{xcolor}
\usepackage{float}
\usepackage{makecell}
\acmJournal{TOG}

\newcommand{\gcc}[1]{#1}

\begin{document}
\title{Topology-Aware Optimization of Gaussian Primitives for Human-Centric Volumetric Videos}

\author{Yuheng Jiang}
\orcid{0000-0001-8121-0015}
\authornote{Equal contributions.}
\affiliation{%
	\institution{Max Planck Institute for Informatics, Saarland Informatics Campus}
	\city{Saarbrücken}
	\country{Germany}
}
\affiliation{%
	\institution{ShanghaiTech University}
	\city{Shanghai}
	\country{China}
}
\affiliation{%
	\institution{ByteDance Inc.}
	\city{Shanghai}
	\country{China}
}
\email{nowheretrix123@gmail.com}

\author{Chengcheng Guo}
\orcid{0009-0003-2122-7001}
\authornotemark[1]
\affiliation{%
	\institution{ShanghaiTech University}
	\city{Shanghai}
	\country{China}
}
\email{guochch2024@shanghaitech.edu.cn}

\author{Yize Wu}
\orcid{0009-0004-0131-114X}
\affiliation{%
	\institution{ShanghaiTech University}
	\city{Shanghai}
	\country{China}
}
\email{wuyize25@163.com}

\author{Yu Hong}
\orcid{0009-0004-1831-7652}
\affiliation{%
	\institution{ShanghaiTech University}
	\city{Shanghai}
	\country{China}
}
\email{hongyu2025@shanghaitech.edu.cn}

\author{Shengkun Zhu}
\orcid{0009-0000-6784-4919}
\affiliation{%
	\institution{ShanghaiTech University}
	\city{Shanghai}
	\country{China}
}
\email{zhushk2024@shanghaitech.edu.cn}

\author{Zhehao Shen}
\orcid{0009-0001-8933-0385}
\affiliation{%
	\institution{ShanghaiTech University}
	\city{Shanghai}
	\country{China}
}
\email{shenzhh2025@shanghaitech.edu.cn}

\author{Yingliang Zhang}
\orcid{0000-0002-0594-7549}
\affiliation{%
	\institution{DGene Digital Technology Co., Ltd.}
	\country{China}
}
\email{yingliang.zhang@dgene.com}

\author{Shaohui Jiao}
\orcid{0009-0002-5350-6809}
\affiliation{%
	\institution{ByteDance Inc.}
	\city{Shanghai}
	\country{China}
}
\email{jiaoshaohui@bytedance.com}

\author{Zhuo Su}
\orcid{0000-0002-7728-0835}
\affiliation{%
	\institution{ByteDance Inc.}
	\city{Shanghai}
	\country{China}
}
\email{suzhuo13@gmail.com}

\author{Lan Xu}
\orcid{0000-0002-8807-7787}
\affiliation{%
	\institution{ShanghaiTech University}
	\city{Shanghai}
	\country{China}
}
\authornote{Corresponding author.}
\email{xulan1@shanghaitech.edu.cn}

\author{Marc Habermann}
\orcid{0000-0003-3899-7515}
\affiliation{%
	\institution{Max Planck Institute for Informatics, Saarland Informatics Campus}
	\city{Saarbrücken}
	\country{Germany}
}
\email{mhaberma@mpi-inf.mpg.de}
\authornotemark[2]

\author{Christian Theobalt}
\orcid{0000-0001-6104-6625}
\affiliation{%
	\institution{Max Planck Institute for Informatics, Saarland Informatics Campus}
	\city{Saarbrücken}
	\country{Germany}
}
\email{theobalt@mpi-inf.mpg.de}
\authornotemark[2]

\begin{abstract}

Volumetric video is emerging as a key medium for digitizing the dynamic physical world, creating the virtual environments with six degrees of freedom to deliver immersive user experiences.
However, robustly modeling general dynamic scenes, especially those involving topological changes while maintaining long-term tracking remains a fundamental challenge.
In this paper, we present \abbtitle, a novel topology-aware dynamic Gaussian representation that disentangles motion and appearance to support, both, long-range tracking and topological adaptation.
We represent scene motion with a sparse set of motion Gaussians, which are continuously updated by a spatio-temporal tracker and photometric cues that detect structural variations across frames.
To capture fine-grained texture, each motion Gaussian anchors and dynamically activates a set of local appearance Gaussians, which are non-rigidly warped to the current frame to provide strong initialization and significantly reduce training time.
This activation mechanism enables efficient modeling of detailed textures and maintains temporal coherence, allowing high-fidelity rendering even under challenging scenarios such as changing clothes.
To enable seamless integration into codec-based volumetric formats, we introduce a global Gaussian Lookup Table that records the lifespan of each Gaussian and organizes attributes into a lifespan-aware 2D layout. This structure aligns naturally with standard video codecs and supports up to 40× compression.
\abbtitle provides a unified, adaptive solution for scalable volumetric video under topological variation, capturing moments where ``elegance in motion'' and ``Power in Stillness''— delivering immersive experiences that harmonize with the physical world.

\end{abstract}

\begin{teaserfigure}
\vspace{-0.3cm}
\raggedright %
{\large \textcolor{magenta}{\texttt{\href{https://guochch.github.io/TaoGS/}{https://guochch.github.io/TaoGS/}}}}\\[4pt]
    \centering
  \includegraphics[width=\textwidth]{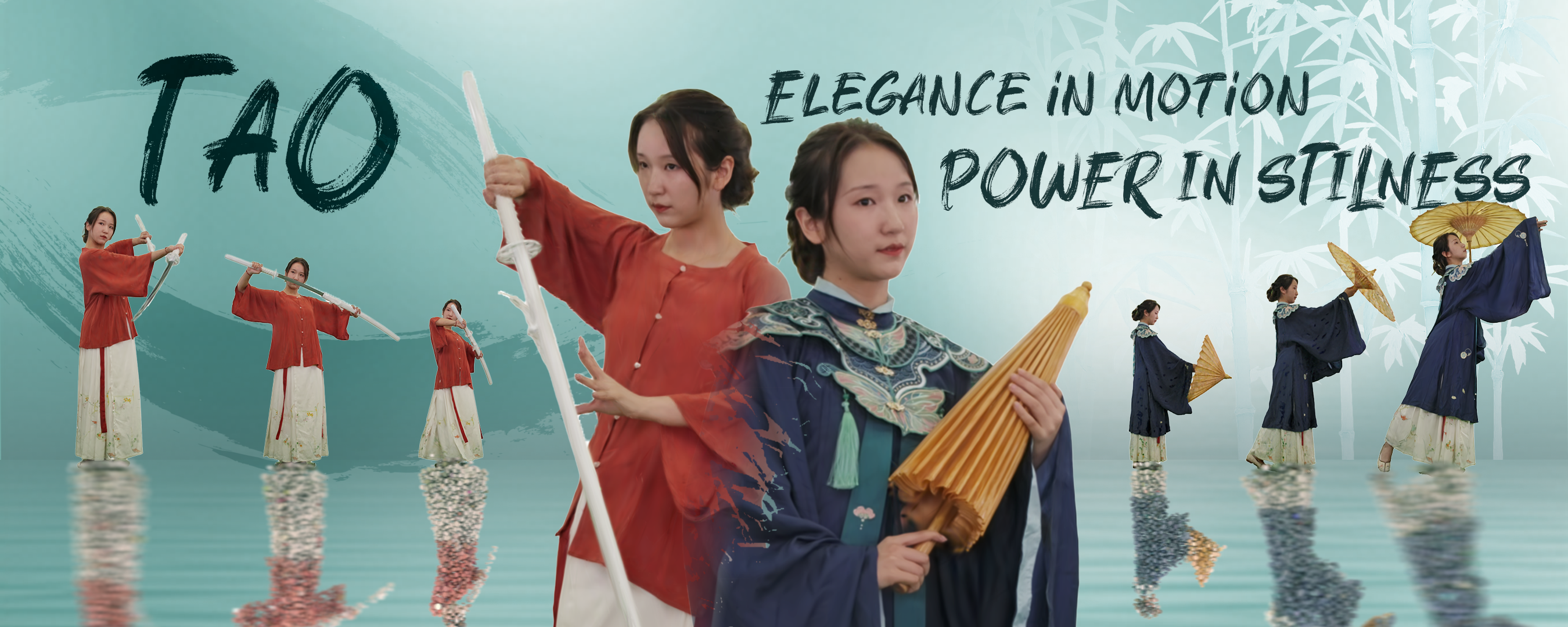}
  \vspace{-20pt}  
  \caption{Elegance in Motion, Power in Stillness. A woman draws her sword beneath still skies, or unfolds her umbrella in silence — moments where grace and power coexist. \abbtitle faithfully captures such long-range, topologically evolving human performances, from subtle motions to complex object interactions.}
  \label{fig:teaser}
\end{teaserfigure}



\maketitle

\section{Introduction}

An increasing portion of the physical world—objects, scenes, cities, and even humans—is gradually entering the process of digitization. As a crucial medium in this transformation, volumetric videos enable the creation of virtual environments with six degrees of viewpoint control. 
The ultimate goal is to create digital replicas of general 4D scenes that faithfully capture both geometry and appearance, and can be rendered in real time with the highest fidelity to deliver immersive user experiences.

This ambition has given rise to two dominant paradigms in volumetric modeling. The first~\cite{yang2023real, duan20244d, li2023spacetime} leverages time-conditioned encoding to model time-evolving appearance, but remains limited to short-range motion.
The second~\cite{luiten2023dynamic, jiang2024robust} focuses on temporally consistent tracking, yet struggles to handle topological changes due to strong regularization constraints.
Both categories have evolved significantly from early textured meshes, to implicit representations, and to the very recent 3D Gaussian splatting~\cite{kerbl20233d}, making high-quality modeling increasingly feasible.

However, a particular overlooked challenge -- and the central focus of this work -- is the long-range tracking and handling of general dynamic scenes involving topological changes, such as a person taking off a jacket. These topological variations and frequent human-object interactions are pervasive in real-world scenarios and cannot be reduced to fixed-topology or human-only assumptions. As a result, despite recent advances in dynamic Gaussian-based methods that achieve high visual fidelity, this problem remains largely unresolved. 
A recent work~\cite{zheng2025gstar} attempts to address this issue by relying on depth input and traditional surface remeshing schemes. However, this approach produces over-smoothed textures and incurs high training costs and memory overhead. Overall, it remains non-trivial to achieve long-term tracking and high-fidelity rendering under topological variations and to model appearance, both, effectively and memory efficiently. 

To tackle this challenge, we propose a new dynamic Gaussian representation, dubbed \abbtitle, which adaptively handles the emergence of new observations and the disappearance of outdated ones. Our method supports robust tracking and topological adaptation, while remaining training-efficient and compression-friendly.
The core of our idea is to leverage a sparse set of topology-aware Gaussians to represent the underlying scene motion, guided by a spatio-temporal tracker and photometric cues, to capture newly emerging observations and continuously update the local deformation graph. Throughout their lifespan, each motion Gaussian derives and activates multiple Gaussians to model fine-grained visual details.

Our method \abbtitle begins with a mesh-free initialization. Leveraging dense matching and triangulation in the first frame, we efficiently initialize a sparse number of Gaussians.
We then perform sequential optimization for robust tracking and topological adaptation. For tracking, we estimate Gaussian motions using an as-rigid-as-possible (ARAP) regularizer to capture non-rigid scene dynamics. To handle topology changes, we employ a spatio-temporal tracker to generate candidate Gaussians that identify structural variations across frames. 
Photometric cues are further incorporated to filter out occluded yet well-reconstructed regions, preventing false detections.
The surviving candidates are then integrated into the training and used to update the local deformation graph. Due to the sparse number of motion Gaussians, the optimization converges quickly. Afterwards, we construct a global lookup table (GLUT) to record the lifespan of each motion Gaussian.

During its lifespan, each motion Gaussian serves as an anchor, deriving and activating a set of Gaussians, termed appearance Gaussians, to express detailed texture. Specifically, we derive new Gaussians based on the local topology of the deformation graph where each motion Gaussian resides, inheriting attributes from it. These appearance Gaussians are activated and non-rigidly warped by their anchors, requiring only minimal fine-tuning to converge and achieve high-fidelity rendering.
During this phase of training, we additionally construct a motion-to-appearance graph to enforce temporal consistency.

As a result, \abbtitle generates topology-aware Gaussians that not only support high-fidelity rendering under topological changes but also naturally align with standard video codec-based volumetric formats. For persistent Gaussians in the GLUT, we apply Morton encoding~\cite{morton1966computer} to project them onto a 2D grid while preserving spatial consistency, which benefits intra-frame prediction in video codecs. For transient Gaussians, we sort them chronologically by activation time to align with inter-frame prediction.
\abbtitle provides a unified, adaptive solution for scalable volumetric video under topological variation, capturing moments of ``Elegance in Motion'' and ``Power in Stillness''— delivering immersive experiences that harmonize with the physical world.

To summarize, our main contributions include:
\begin{itemize} 
	\setlength\itemsep{0em}

	\item We introduce a topology-aware Gaussian representation that supports long-term tracking and high-fidelity rendering under topological variations.
    
	\item We propose a spatio-temporal tracking strategy to adaptively detect topological changes and efficiently generate Gaussians in the corresponding regions.
    
	\item We adopt an activation mechanism on top of a motion-to-appearance Gaussian representation that efficiently models scene details and enables fast training.
	
	\item Our representation is inherently compatible with standard video codec-based volumetric formats, achieving up to 40× compression even under complex topological changes.
\end{itemize}

\section{Related Work} 
\paragraph{Non-rigid Reconstruction.}
Non-rigid reconstruction has been extensively explored in the context of human performance capture~\cite{zollhofer2018state, newcombe2015dynamicfusion, guo2017real, slavcheva2018sobolevfusion,xu2019deep, xiang2020monoclothcap,  shao2022floren,realTimeDDC,kwon2024deliffas}. 
Early approaches~\cite{taylor2012vitruvian, DoubleFusion, xu2019deep, UnstructureLan, robustfusion, su2022robustfusionPlus} typically rely on full body template fitting or human parametric model~\cite{SMPL2015}.
Recent neural advances, represented by NeRF~\cite{nerf}, have inspired a series of methods~\cite{park2021nerfies,  fang2022fast, habermann2023hdhumans, zhu2023trihuman, jiang2022neuralhofusion, jiang2023instant, habermann2021real} for dynamic reconstruction based on volumetric neural representations. 
3DGS~\cite{kerbl20233d} embraces explicit representations to enable real-time and high-quality scene modeling. Several dynamic variants~\cite{wu20234d, xu20234k4d, yang2023deformable, jena2023splatarmor,moreau2024human, qian20243dgs,pan2024soar,wang2024v} leverage MLPs to model temporal correspondences across frames. 
GaussianAvatars~\cite{qian2023gaussianavatars, chen2023monogaussianavatar} bind Gaussians to the FLAME mesh for articulated motion, while D3GA~\cite{zielonka2023drivable} utilizes tetrahedral cages for deformation control. \gcc{Several methods~\cite{huang2024sc, scaffoldgs, jiang2025reperformer, lei2025mosca} adopt a layer-wise representation to disentangle motion and appearance}. However, these techniques suffer from slow training and rely on fixed topology templates, making them unsuitable for handling topological changes.

\paragraph{Topology-Aware Dynamic Representations.} 
Handling topological variations remains a significant challenge in dynamic scene representation. Existing topology-aware approaches typically follow two main strategies. The first leverages keyframe-based mechanisms~\cite{collet2015high,dou2016fusion4d, motion2fusion}, dividing the temporal sequence into discrete segments to mitigate the effects of topological changes.
HiFi4G~\cite{jiang2023hifi4g} and $V^3$~\cite{wang2024v} further integrate this keyframe-based design with Gaussian splatting, enabling high-fidelity rendering under challenging topological changes. 
Another strategy uses time-conditioned encoding to jointly represent motion and appearance within a unified framework. Several NeRF-based methods~\cite{park2021hypernerf, zheng2023editablenerf, isik2023humanrf, fridovich2023k, cao2023hexplane, shao2023tensor4d} adopt this idea by learning 4D feature grids or decomposed tensor volumes that encode space-time variations. Recent Gaussian-based methods~\cite{li2023spacetime, xu20234k4d, duan:2024:4drotorgs, xu2024longvolcap, yang2023real} extend 3DGS to the spatio-temporal domain by embedding motion and appearance into a unified 4D Gaussian representation. 
Yet, these approaches often struggle with unreliable tracking, which limits their ability to maintain temporal coherence and structural consistency, key factors for real-world dynamic scene applications.
More recently, GauSTAR~\cite{zheng2025gstar} attempts to address the tracking and topological 
\gcc{adaptation} by introducing an explicit surface-based representation with remeshing to handle topological changes.
However, GauSTAR tends to produce over-smoothed rendering results and suffers from long training times and high memory consumption.
In contrast, our method leverages a compact Gaussian Lookup Table and activatable appearance Gaussians to significantly reduce training time and storage overhead, while achieving high-fidelity rendering under topological variations.

\paragraph{Volumetric Compression.}
Compact representations are essential for efficient 3D/4D reconstruction and have been extensively explored in both traditional and modern neural approaches. For mesh compression, early works aim to reduce geometric redundancy through techniques such as PCA-based compression~\cite{alexa2000representing,vasa2007coddyac,luo2013compression}, mesh segmentation~\cite{gupta2002compression, mamou2009tfan}, and vertex trajectory prediction~\cite{mamou2009tfan, luo2013compression}, which help maintain structural consistency and temporal coherence.
In the era of implicit representations, compactness has been pursued through CP decomposition~\cite{chen2022tensorf}, low-rank approximation~\cite{tang2022compressible}, neural codecs~\cite{Wang_2023_CVPR}, and tri-plane factorization~\cite{reiser2023merf, hu2023tri}, where 3D fields are represented using multiple 2D feature planes for efficient storage and rendering.
More recently, there has been growing interest in compressing 3D Gaussian-based representations. Approaches include vector quantization and entropy coding~\cite{navaneet2023compact3d, Niedermayr_2024_CVPR, Lee_2024_CVPR}, anchor-based sparsification~\cite{scaffoldgs}, and hash table~\cite{chen2024hac} all aimed at reducing storage while preserving fidelity.
To further support compression in dynamic scenes,  4K4D~\cite{xu20234k4d} proposes a unified 4D feature grid, while DualGS~\cite{jiang2024robust} adopts a layer-wise decomposition and TeTriRF~\cite{wu2024tetrirf} transforms the density grid into tri-plane representations. VideoRF~\cite{wang2023videorf} encodes 4D radiance fields as 2D feature streams. HiFi4G~\cite{jiang2023hifi4g} and V$^3$~\cite{wang2024v} 
\gcc{extend} compression with keyframe setting to Gaussian splatting, enabling streamable and compact Gaussian-based rendering. In contrast, our method handles topological changes while avoiding the costly keyframe resetting process. It seamlessly packs into a 2D Gaussian attribute map for compression, a capability not demonstrated by previous methods.

\begin{figure*}[t] 
	\begin{center} 
		\includegraphics[width=\linewidth]{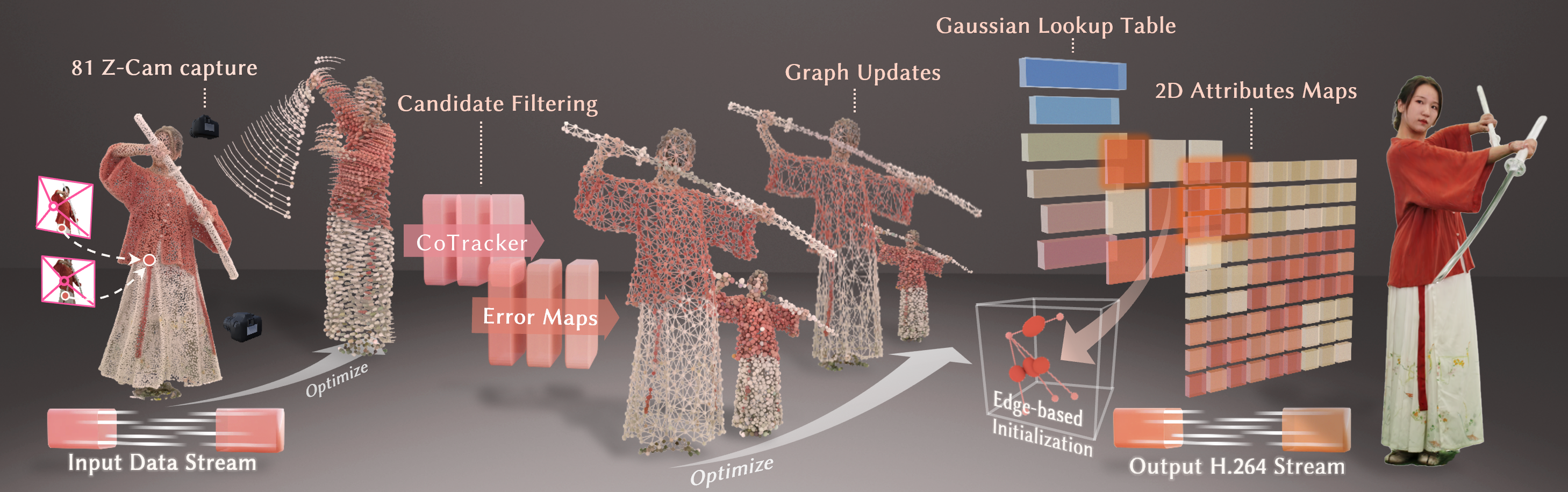} 
	\end{center} 
    \vspace{-10pt}  
    \caption{We propose a novel motion-to-appearance Gaussian representation for robust tracking and high-fidelity rendering of general 4D scenes with topological changes. We track sparse motion Gaussians and incorporate new candidate Gaussians through a spatial-temporal tracker and error map to model new observations. The motion Gaussians are then transformed into a Gaussian Look-Up Table (GLUT), activating corresponding appearance Gaussians, which can be packed into 2D attribute maps for efficient video codec compression.}
    \label{fig:pipeline} 
    \vspace{-8pt}
\end{figure*}

\section{\abbtitle}\label{sec:overview} 
To handle complex topological changes in multi-view performance capture, we employ a dynamic Gaussian representation that enables both efficient compression and high-quality rendering. The methodology, visualized in Fig.~\ref{fig:pipeline}, consists of two main components. We first introduce topology-aware motion Gaussians along with a local rigidity regularizer to capture underlying scene dynamics. Leveraging spatio-temporal trackers, we adaptively insert candidate Gaussians to account for newly emerging or disappearing observations, while updating the local deformation graph. 
The second component derives a set of $K$ activatable appearance Gaussians from each motion Gaussian to represent fine-grained texture. These appearance Gaussians are triggered by their anchored motion Gaussians and are compactly stored in 2D attribute maps for further compression~\cite{wang2023videorf, wang2024v}. 
This decoupled representation reduces redundancy and accelerates training by isolating motion and appearance modeling. Our method organically balances long-range robust tracking with topological adaptability, achieving real-time high-fidelity rendering even under a 40× compression ratio.

\subsection{Topology-aware Motion Registration} \label{sec:motion}
To enable long-range tracking and topological adaptation, we initialize motion Gaussians in the first frame using known camera parameters, followed by topological detection and graph updates via frame-by-frame registration.
\paragraph{First Frame Initialization.} 
Instead of relying on SfM~\cite{7780814} or mesh initialization, which often misses fine structures like hair or blade edges, we follow EDGS~\cite{kotovenko2025edgs} and triangulate dense 2D correspondences for initialization. We first group all input cameras into nearest-neighbor pairs and utilize a pretrained dense matcher~\cite{edstedt2024roma} to extract dense pixel-wise correspondences, which are triangulated via least squares to produce an accurate point cloud. 
We then perform training from this point cloud and prune it by opacity, retaining approximately 20,000 Gaussians $\mathcal{G}$ characterized by position $p$, rotation $q$, third-order spherical 
\gcc{harmonics} $\mathcal{C}$, opacity $\sigma$, and scaling $s$.
We jointly optimize these Gaussians using the following loss:
\begin{equation}
E_{\text{init}} = \lambda_{\text{lap}} E_{\text{lap}} + \lambda_{\text{iso}} E_{\text{iso}} + \lambda_{\text{size}} E_{\text{size}} + E_{\text{color}},
\end{equation}
where \( E_{\text{lap}} \) denotes the Laplacian smoothness loss, which promotes a non-overlapping, spatially regular distribution that approximates a zero-level set of a surface, facilitating the following optimization:
\begin{equation}
\begin{split}
E_{\text{lap}} = \frac{1}{N} \sum_{i=1}^{N} \left\| p_{i,1} - sg\left[ \frac{1}{K} \sum_{j \in \mathcal{N}(i)} (p_{j,1}) \right] \right\|^2,
\end{split}
\end{equation}
where \( \text{sg}[\cdot] \) denotes the stop-gradient operator, \( p_{i,1} \) is the \( i \)-th Gaussian position at the first frame, and \(\mathcal{N}(i) \) denotes its k-nearest neighbors(KNN).
The color loss \( E_{\text{color}} \) follows the 3DGS. For isotropic loss \( E_{\text{iso}} \) and the size loss \( E_{\text{size}} \),  we refer to DualGS~\cite{jiang2024robust}.

\begin{figure}[tbp] 
	\centering 
	\includegraphics[width=1\linewidth]{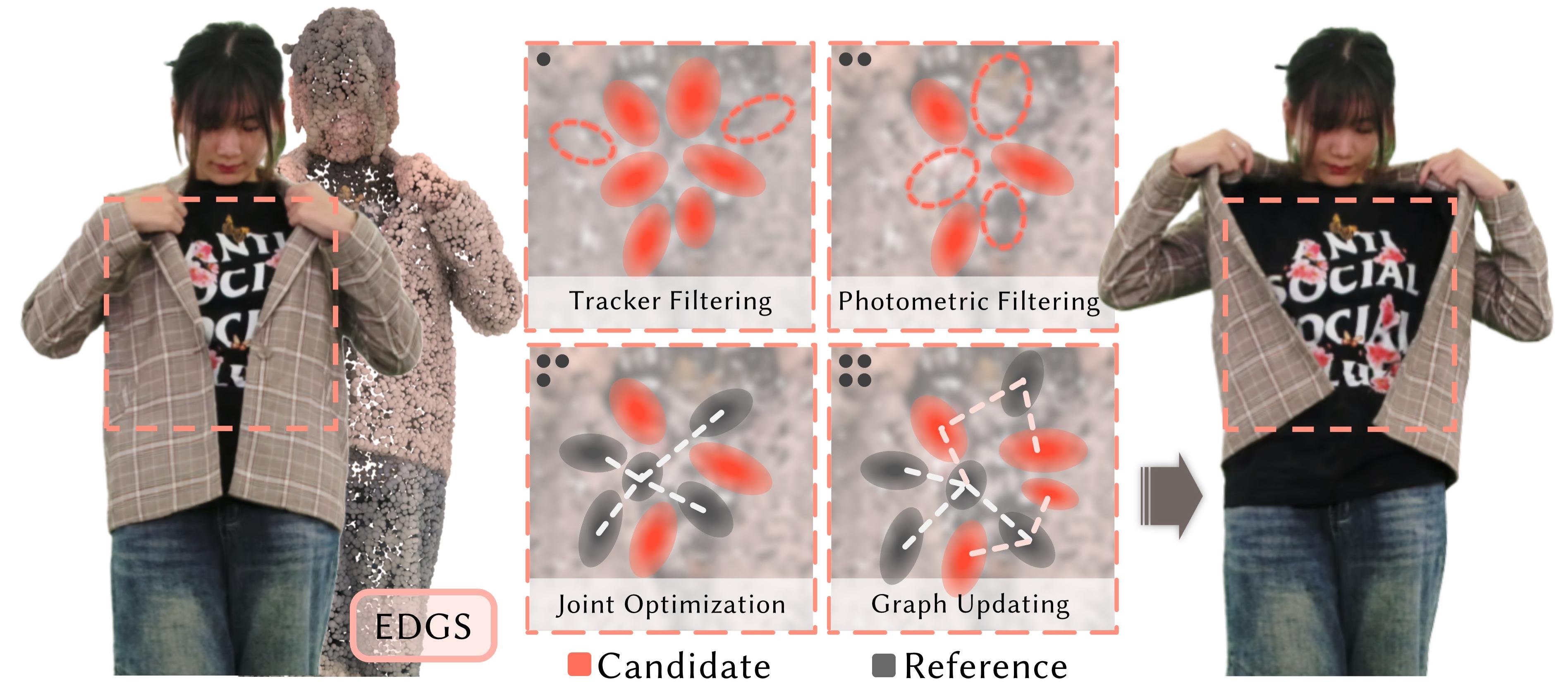} 
	\vspace{-15pt} 
	\caption{   
    Illustration of our Candidate Filtering and graph updating strategy. \gcc{
    (1) We employ a spatio-temporal tracker to filter the candidate Gaussians.
    (2) We further incorporate photometric error to identify poorly reconstructed regions in $\mathcal{G}_\mathrm{cand}$.
    (3) The remaining $\mathcal{G}_\mathrm{cand}$ are jointly optimized with $\mathcal{G}_\mathrm{ref}$.
    (4) During optimization, we periodically update the graph to preserve these connections.}}
	\label{fig:detection} 
	\vspace{-16pt} 
\end{figure}

After initialization, we proceed with dynamic optimization.
To enable long-term tracking, inspired by Dynamic3DGaussian~\cite{luiten2023dynamic}, we introduce non-rigid physical modeling to regularize optimization and ensure physically plausible motion. Specifically, we connect each Gaussian to its KNN to form a deformation graph and employ an as-rigid-as-possible (ARAP) constraint to regularize both position $p$ and rotation $q$: 
\begin{equation}
\begin{aligned}   
E_{\text {smooth }}= & \sum_i \sum_{k \in \mathcal{N}(i)} w_{i, k}^{t-1} \| R\left(q_{i, t} * {q_{i, t-1}}^{-1}\right) \\
& \left(p_{k, t-1}-p_{i, t-1}\right)-\left(p_{k, t}-p_{i, t}\right) \|_2^2,
\end{aligned}
\end{equation}
where $R(\cdot)$ converts a quaternion to a rotation matrix and $w$ corresponds to the blending weights 
\(
w_{i,k}^{t}=\exp \left(-\left\|p_{i,t}-p_{k,t}\right\|_2^2 / l^2\right) \).
$l$ is the influence radius.
The optimization equation is then defined as:
\begin{equation}
E_{\text{}} = \lambda_{\text{smooth}} E_{\text{smooth}} + E_{\text{color}},
\end{equation}
We refer to these Gaussians, whose motion follows physically plausible constraints, as motion Gaussians. However, such Gaussians are inherently incapable of handling topological changes, let alone accounting for newly emerging observations. To address this, as shown in Fig.~\ref{fig:detection}, we extend the ARAP regularizer with densification guided by spatio-temporal trackers to detect structural variations, insert new candidate Gaussians $\mathcal{G}_\mathrm{cand}$ into reference motion Gaussians set $\mathcal{G}_\mathrm{ref}$, and update the graph during optimization to preserve local rigidity. We elaborate on each of these steps below.
\paragraph{Candidate Filtering.} Analogous to the initialization, we efficiently generate candidate Gaussians for frame $t$ from paired images using dense spatial matching. However, EDGS~\cite{kotovenko2025edgs} yields an overly dense candidate pool, with many Gaussians corresponding to topologically stable or previously well-reconstructed regions. To reduce redundancy, we retain only those associated with new topological changes between adjacent frames, as identified by a temporal tracker and photometric cues.
To this end, for each view $v$, we identify regions newly observed in frame $t$ relative to frame $t-1$. We apply a temporal tracker~\cite{karaev2024cotracker3} to estimate the pixel-wise optical flow correspondences $\mathcal{T}$, formalized as:
\begin{equation}
\mathcal{T}(I^v_t, I^v_{t-1}) \rightarrow \left( \mathcal{W}_{t \rightarrow t-1}^v, \mathcal{M}_{t \rightarrow t-1}^v \right),
\end{equation}
where \( \mathcal{W}_{t \rightarrow t-1}^v \in \mathbb{R}^{2 \times H \times W} \) is a dense warp field that maps each pixel in \( I^v_t \) to its corresponding location in \( I^v_{t-1} \), and \( \mathcal{M}_{t \rightarrow t-1}^v \in \{0,1\}^{H \times W} \) is a binary mask, where 0 indicates a newly observed pixel with no correspondence in $I^v_{t-1}$. We leverage this mask to filter candidate Gaussians via:
\begin{equation}
\begin{aligned}
w_1(\mathcal{G}_{cand}^k) &= (1-\mathcal{M}_{t \rightarrow t-1}^{v_1}(u_1)) \times (1-\mathcal{M}_{t \rightarrow t-1}^{v_2}(u_2)),
\end{aligned}
\end{equation}
where $u_1$ and $u_2$ are the corresponding pixels in the paired views $v_1$ and $v_2$, identified by the spatial tracker and triangulated to construct $\mathcal{G}_{cand}^k$. We retain only candidates with $w_1(\mathcal{G}_{cand}^k) = 1$.
This strategy significantly reduces the number of candidates. However, when pixel-level correspondences are missing, the temporal tracker alone cannot reliably distinguish between occlusions and genuinely new observations. Since occluded regions are already well reconstructed, misidentifying them as topological changes would introduce redundant Gaussians and degrade representation efficiency.
To address this ambiguity, we incorporate photometric error as an auxiliary filter to identify poorly reconstructed regions, further suppressing occlusion-induced false positives. The overall filtering is defined as:
\begin{equation}
\begin{aligned}
w_2(\mathcal{G}_{cand}^k) &= E_{\text{color}}^{v_1}(u_1) + E_{\text{color}}^{v_2}(u_2).
\end{aligned}
\end{equation}
Leveraging the correspondences between 3D candidates and image pixels established by the spatial tracker, we discard the candidate when the confidence score $w_2(\mathcal{G}_{cand})$ falls below a threshold $\epsilon$.

\paragraph{Graph updates.}
The remaining Gaussians $\mathcal{G}_\mathrm{cand}$ are jointly optimized with the reference set $\mathcal{G}_\mathrm{ref}$. During optimization, candidates are cloned or split, yielding new Gaussians that are incorporated into the pool. To maintain coherent motion and prevent drift, we update the deformation graph by connecting each newly generated Gaussian to its KNN among all Gaussians.
Besides, Gaussians with opacity below a predefined threshold are removed along with their associated graph connections.
We also periodically update the local graph to balance two objectives, preventing the drift of $\mathcal{G}_\mathrm{cand}$ and preserving the connection of the $\mathcal{G}_\mathrm{ref}$ subgraph. This is achieved by marking each Gaussian in $\mathcal{G}_\mathrm{cand}$ and its 2-ring neighbors, then recomputing their nearest neighbors to refresh local connectivity. 

\noindent{\bf Implementation.}
For the initialization, we use the following empirical weights:
$\lambda_{\text{lap}}=2.0, \lambda_{\text{iso}}=0.001,\lambda_{\text{size}}=1.0$. 
We train each frame for 6,000 iterations. 
During the motion registration phase, we set $\lambda_{\text{smooth}}=0.05$. The first 3000 iterations capture non-rigid motions. At iteration 3000, we insert candidate Gaussians proposed by the spatial-temporal tracker. 
\gcc{Every 300 iterations, we perform densification and pruning, as well as update only the local KNN graph of approximately 20,000 motion Gaussians, which takes $2.40$s per frame, incurring negligible overhead.} For candidate Gaussians, we optimize all attribute parameters, whereas for reference motion Gaussians, we optimize only position, rotation, and opacity. After training each frame, all surviving candidate Gaussians $\mathcal{G}_\mathrm{cand}$ are merged into the reference set $\mathcal{G}_\mathrm{ref}$, and the KNN graph is recomputed accordingly. To detect topological changes, we downsample the image to $324 \times 576$ and utilize CoTracker~\cite{karaev2024cotracker3} to estimate optical flow, and set $\epsilon=0.46$.

\vspace{-12pt}
\begin{figure*}[ht!] 
	\begin{center} 
		\includegraphics[width=0.99\linewidth]{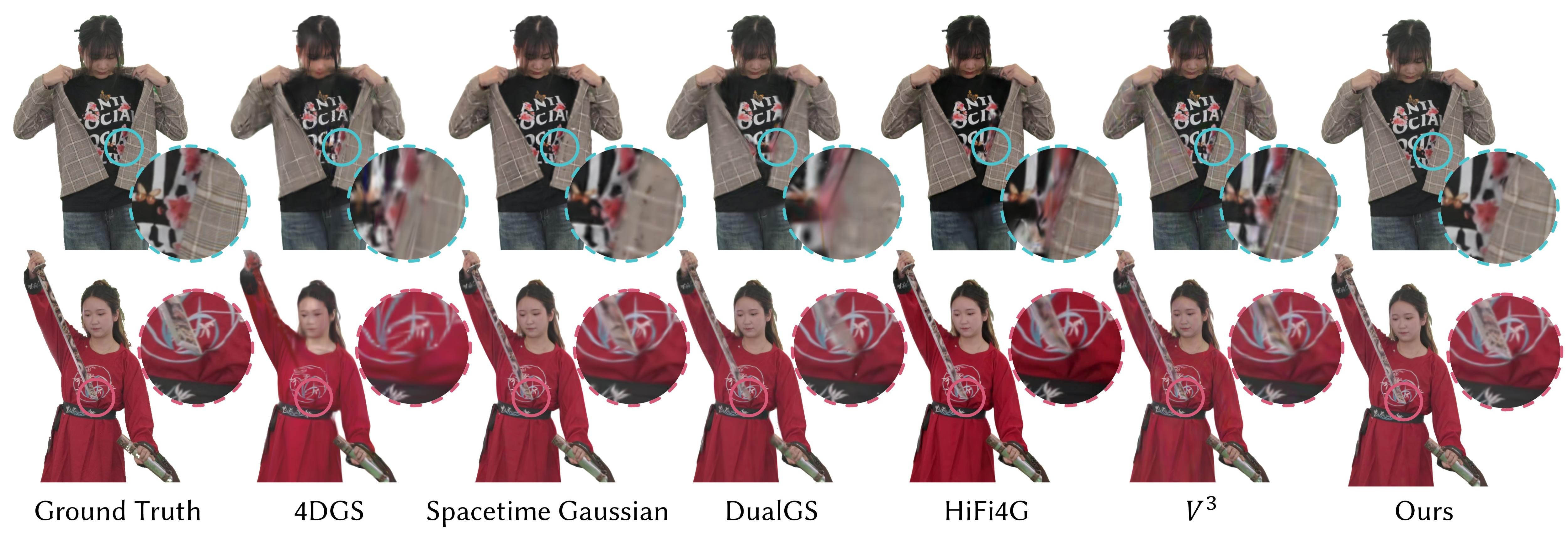} 
	\end{center} 
	\vspace{-13pt}
	\caption{Qualitative comparison of our method against 4DGS~\cite{yang2023gs4d}, Spacetime Gaussian~\cite{li2023spacetime}, DualGS~\cite{jiang2024robust}, HiFi4G~\cite{jiang2023hifi4g} and V$^3$~\cite{wang2024v} on our challenging dataset. Our method achieves the highest rendering quality.}
	\label{fig:fig_comp_1}
 	\vspace{-10pt}
\end{figure*} 

\subsection{Activatable Appearance Gaussians} \label{sec:appearance}

While the sparse motion Gaussians effectively capture the scene dynamics, they are insufficient for high-fidelity rendering. To model fine-grained texture, we introduce an auxiliary set of denser Gaussians, termed appearance Gaussians, which are non-rigidly warped via motion Gaussians to the current frame, providing strong initialization and significantly reducing training time. However, accommodating topological changes still necessitates the insertion of appearance Gaussians to reflect newly observed regions, which introduces notable overhead in both training and compression.

To mitigate this, we propose a lightweight activation mechanism where sparse motion Gaussians govern the activation of dense appearance Gaussians. New appearance Gaussians are sampled and activated upon motion Gaussian insertion, and deactivated upon pruning. This motion-level topological control reduces the complexity of directly managing topological adaptation at the appearance level.
Nevertheless, the frequent variation in the number of active Gaussians severely impairs compression efficiency. To address this, we introduce a global Gaussian Lookup Table (GLUT) that records the lifespan of each motion Gaussian and maps frame-specific Gaussians to a unified index space, enabling consistent cross-frame representation and forming the foundation of our compression scheme.

\paragraph{Edge-based Initialization.} 
We efficiently initialize each appearance Gaussian at its first activation by inheriting attributes from its anchored motion Gaussian. 
To derive these Gaussians, we re-interpret the KNN graph, where each Gaussian $\mathcal{G}^t_i$ serves as a node connected to its $K$ nearest spatial neighbors, forming edges that define a local graph structure. 
For each motion Gaussian, we sample one appearance Gaussian along each of its K outgoing edges, yielding a consistent number of K appearance Gaussians for structured compression.
The position of each appearance Gaussian is sampled along its respective edge, biased between one-third and one-half of the edge length toward the anchor $\mathcal{G}^t_i$. Its scaling is determined by the distance to the anchor, while its rotation, color, and opacity are directly inherited to ensure coherent and consistent initialization.
At the beginning of each frame’s training, motion Gaussians within their lifespan activate and non-rigidly deform their corresponding appearance Gaussians, providing strong initialization for faster convergence.
During training, we further enforce local rigidity by linking each appearance Gaussian to its anchor motion Gaussian, promoting spatial coherence.

\noindent{\bf Implementation.}
We allocate 6,000 iterations per frame to the appearance training phase and optimize all attributes of each appearance Gaussian with $\lambda_{\text{smooth}}=0.0002$. Every motion Gaussian recorded in the GLUT creates and activates 9 appearance Gaussians. We follow the temporal regularizer from HiFi4G~\cite{jiang2023hifi4g} to ensure temporal consistency of Gaussian attributes $\mathcal{C}$, $\sigma$ and $s$.

\subsection{Lifespan-Aware Compression}
\label{sec:compression}

Our topology-aware motion-appearance representation achieves temporal consistency through a global Gaussian Lookup Table (GLUT), which assigns a unique index to each motion Gaussian and records its lifespan across frames. The GLUT is further mapped to a 2D layout, enabling consistent referencing of appearance Gaussians across frames, even during topological changes. Inactive Gaussians inherit attributes from the last active frame to maintain temporal coherence.
This structure integrates seamlessly with standard video codec-based volumetric formats, enhancing compressibility.

We also introduce a lifespan-aware compression strategy to spatially sort motion Gaussians indexed in the GLUT. This ordering is propagated to appearance Gaussians to ensure layout consistency across levels.
Based on their lifespan, Gaussians are classified as either persistent or transient. For persistent Gaussians, we quantize their 3D positions in the first frame and apply Morton ordering~\cite{morton1966computer} by interleaving the binary representations of their spatial coordinates to construct a 2D grid layout. This mapping preserves the spatial locality of neighboring Gaussians in UV space, allowing video encoders to better exploit spatial redundancy during intra-frame prediction.
For transient Gaussians, we arrange them in a row-major order based on their activation timestamps. This layout enhances temporal consistency and predictability across frames, aligning well with the inter-frame prediction mechanism of H.264 encoders, allowing the codec to better capture temporal redundancy.

We further encode the structured 2D attribute maps as image sequences using off-the-shelf video codecs~\cite{wang2024v, wu2024tetrirf}. Even under complex topological changes, our approach achieves a 40× compression ratio while preserving high-fidelity rendering.

\section{Experiments} 

\begin{figure*}[htbp] 
	\begin{center} 
		\includegraphics[width=1\linewidth]{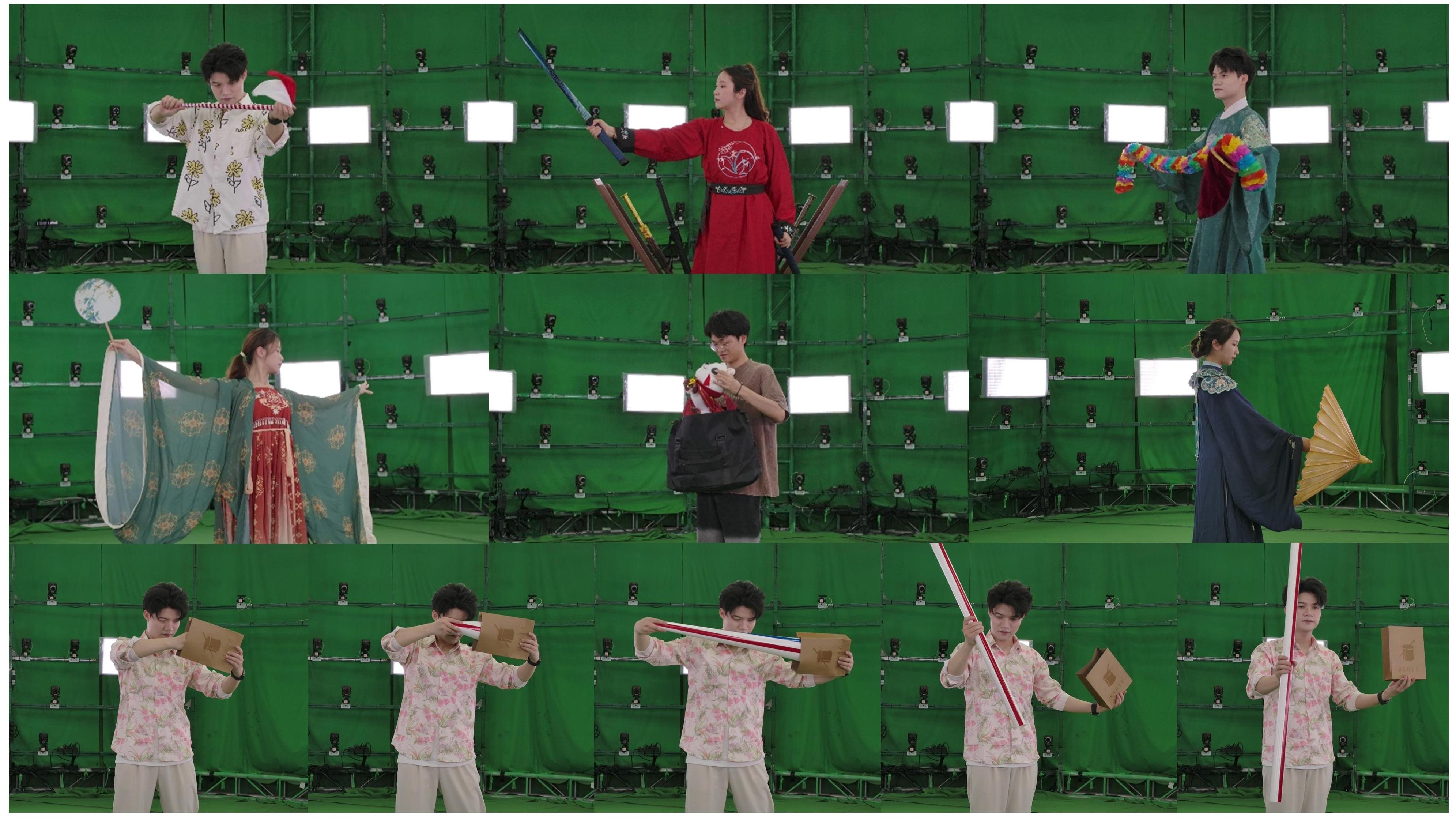} 
	\end{center} 
	\vspace{-10pt}
	\caption{Overview of our dataset. The first two rows show representative static samples from our multi-view capture system. The bottom row displays consecutive frames from a dynamic sequence, illustrating the topological changes we address.} 
	\label{fig:dataset}
	\vspace{-5pt}
\end{figure*}

\begin{figure*}[htbp] 
	\begin{center} 
		\includegraphics[width=1\linewidth]{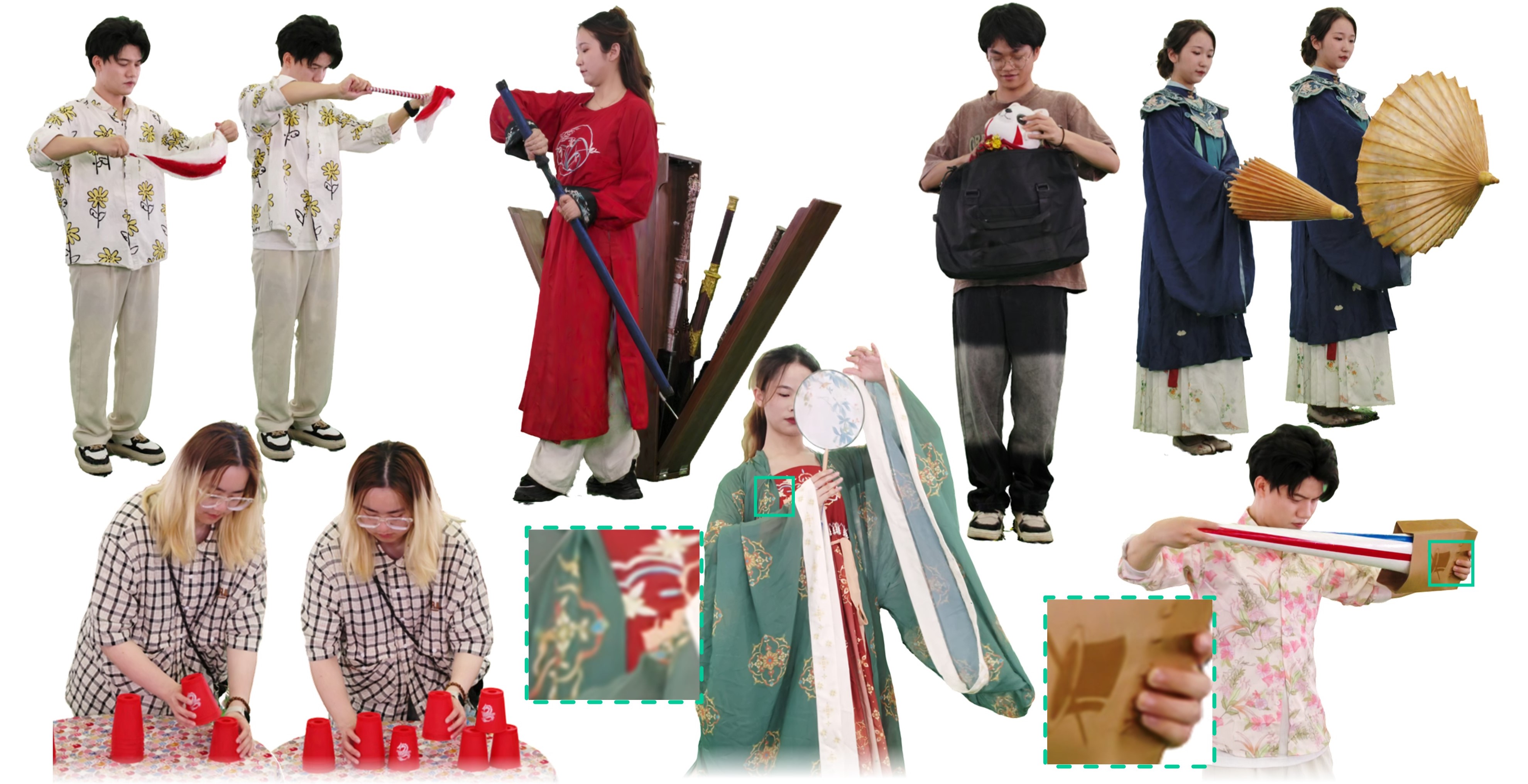} 
	\end{center} 
	\vspace{-10pt}
	\caption{ Gallery of our results under topological changes. \abbtitle achieves robust tracking and high-fidelity rendering while retaining a 40× compression ratio.} 
	\label{fig:gallery}
	\vspace{-5pt}
\end{figure*}

\begin{figure*}[htbp] 
	\begin{center} 
		\includegraphics[width=1.0\linewidth]{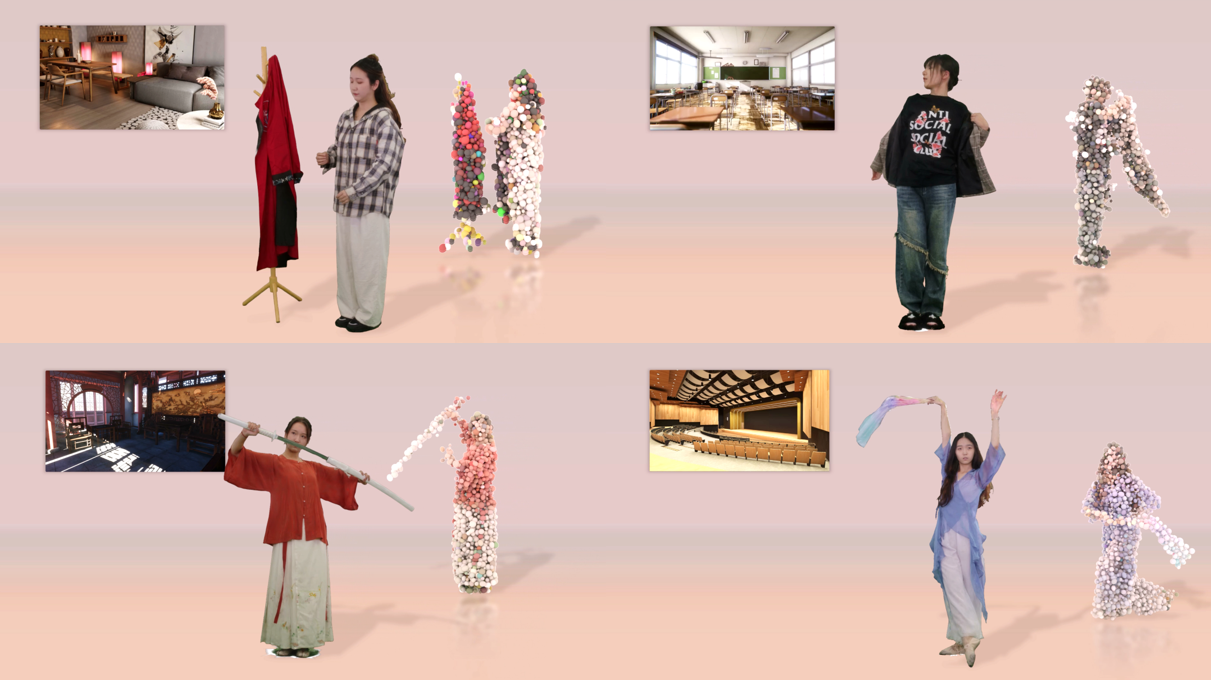} 
	\end{center} 
	\vspace{-5pt}
	\caption{We demonstrate the rendering of various topological change scenarios along with the corresponding motion Gaussians, such as "removing clothes", "drawing a sword", and "fan dance".} 
	\label{fig:app_1}
	\vspace{-5pt}
\end{figure*}

\begin{figure*}[htbp] 
	\begin{center} 
		\includegraphics[width=1.0\linewidth]{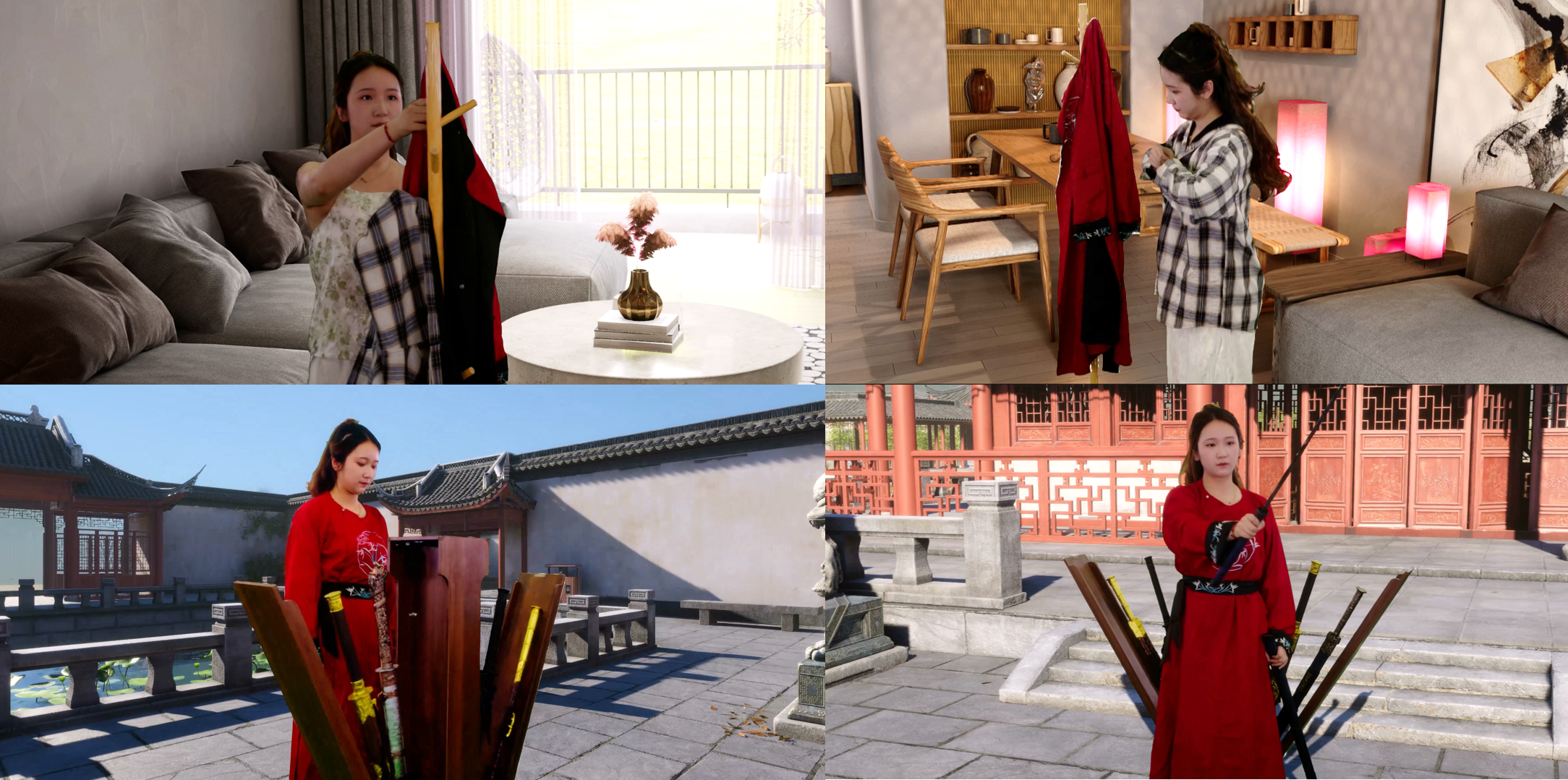} 
	\end{center} 
	\vspace{-5pt}
	\caption{We showcase the seamless integration of 4D assets with static environments, providing users with an immersive experience. For more dynamic details, please refer to our supplementary video.} 
	\label{fig:app_2}
	\vspace{-5pt}
\end{figure*}

To further demonstrate the capabilities of \abbtitle, we conduct human performance capture at 30 FPS from 81 camera views at 4K resolution, covering diverse sequences such as opening a sword box, unsheathing a sword, cup stacking, and magic performances involving production and transformation, \gcc{as illustrated in Fig.~\ref{fig:dataset}.}
Leveraging this dataset, which features a variety of topological changes, our method consistently achieves robust underlying motion tracking and high-fidelity rendering. 
\gcc{Please refer to Appendix Fig.~\ref{fig:gallery} and Fig.~\ref{fig:app_1} for more rendering results, and Fig.~\ref{fig:app_2} for VR applications.}
Built on the original 3DGS~\cite{kerbl20233d}, we integrate Taming 3DGS~\cite{mallick2024taming} for faster training, reaching 2 minutes per frame on an NVIDIA RTX 4090 GPU. 
\gcc{
Furthermore, \abbtitle adopts a train-then-compress pipeline and naturally supports conversion into compact 2D Gaussian streams, which can be transmitted in real time through standard video codecs, enabling immersive volumetric playback on mobile and VR devices.}

\vspace{-10pt}
\subsection{Comparison} 

We compare \abbtitle against prior dynamic Gaussian representation, including time-conditioned methods 4DGS~\cite{yang2023gs4d} and Spacetime Gaussian~\cite{li2023spacetime}, as well as tracking-based methods DualGS~\cite{jiang2024robust}, HiFi4G~\cite{jiang2023hifi4g} and V$^3$~\cite{wang2024v}, using our dataset characterized by significant topological changes.
As shown in Fig.~\ref{fig:fig_comp_1}, 4DGS often fails or produces spiky artifacts, while Spacetime Gaussian struggles with high-frequency details. Both methods generate excessive Gaussians to handle time-varying appearance, lack long-range tracking, and incur high storage costs.
DualGS relies on strong rigidity constraints, causing blurry renderings. HiFi4G and V$^3$ use costly keyframes and suffer from intra-segment distortions.
In contrast, \abbtitle leverages a spatio-temporal tracker to adaptively insert new motion Gaussians and activate associated appearance Gaussians, enabling robust tracking and high-fidelity rendering under topology changes.
For quantitative evaluation, we benchmark all methods on three sequences of 200 frames using PSNR, SSIM, LPIPS, storage, \gcc{training time, and rendering FPS}. As shown in Tab.~\ref{table:comparison1} \gcc{and Tab.~\ref{table:comparison2}}, 
\gcc{\abbtitle not only achieves the highest rendering quality in challenging scenarios, but also exhibits efficient storage footprint, training speed, and real-time rendering performance.}
In terms of storage, DualGS achieves high compression but fails with topological changes. V$^3$ reduces storage by pruning Gaussians and using zeroth-order Spherical Harmonics, at the cost of severely degraded visual fidelity.
In contrast, \abbtitle combines robust, topology-aware modeling with efficient training and compression, requiring only 1.335 MB per frame and achieving a 40× compression ratio.

\begin{figure}[t] 
	\begin{center} 
		\includegraphics[width=1\linewidth]{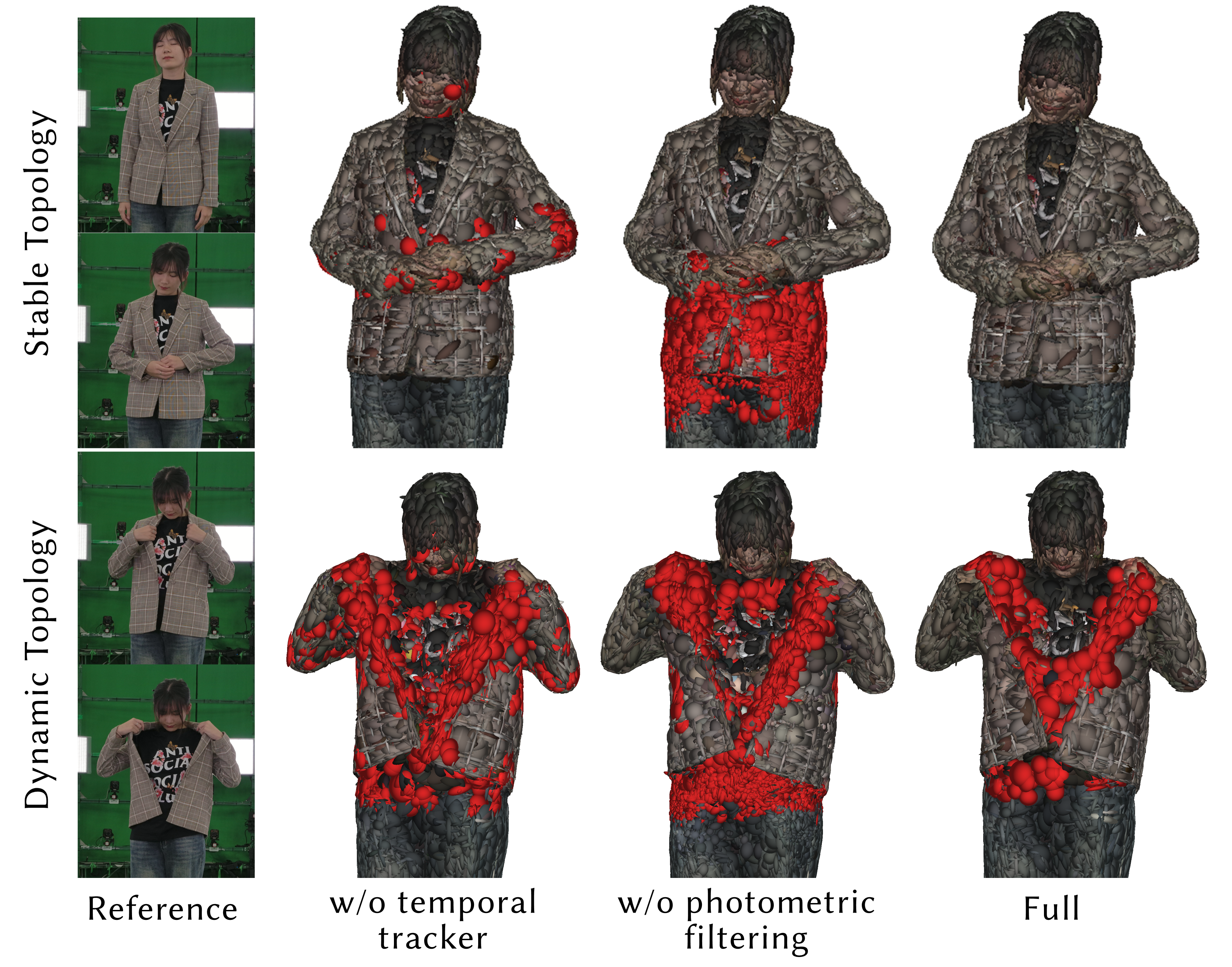}
	\end{center} 
	\vspace{-15pt}
	\caption{Evaluation of topology-aware Gaussian densification strategy.}
	\label{eval:tracker}
	\vspace{-12pt}
\end{figure}

\begin{table}[t]
	\begin{center}
		\centering
		\vspace{-1pt}
        \caption{
       \textbf{\gcc{Rendering quality} comparison on our dataset}. Green, yellow, and red cell colors indicate the best, second-best, and third-best results, respectively.}
        \vspace{-6pt}
		\resizebox{0.47\textwidth}{!}{
			\begin{tabular}{l|cccccc}
				\hline
				Method   &  PSNR $\uparrow$ & SSIM $\uparrow$ & LPIPS $\downarrow$  \\
				\hline
				4DGS~\cite{yang2023gs4d}\qquad\qquad & 30.654 & 0.9644 &  0.0536  \\ 
                Spacetime Gaussian~\cite{li2023spacetime}     & 34.079 & 0.9834 & 0.0256  \\
                DualGS~\cite{jiang2024robust} & 34.092 & 0.9800 & 0.0383\\
				HiFi4G~\cite{jiang2023hifi4g}  & 34.870 & 0.9845 & \colorbox{best3}{0.0246}  \\
				$V^3$~\cite{wang2024v}  & \colorbox{best3}{35.043} & \colorbox{best3}{0.9859} & 0.0296  \\
    		\hline
                Ours(Before Compression)  & \colorbox{best1}{37.264} & \colorbox{best1}{0.9911} & \colorbox{best1}{0.0182} \\
                Ours(After Compression) & \colorbox{best2}{36.697} & \colorbox{best2}{0.9881} & \colorbox{best2}{0.0214}   \\
				\hline
			\end{tabular}
		}
		\label{table:comparison1}
    	\vspace{-10pt}
	\end{center}
\end{table}

\begin{table}[t]
	\begin{center}
		\centering
		\vspace{-1pt}
        \caption{
       \gcc{\textbf{Performance comparison on our dataset}. Green, yellow, and red cell colors indicate the best, second-best, and third-best results, respectively.}}
        \vspace{-6pt}
		\resizebox{0.47\textwidth}{!}{
			\begin{tabular}{l|ccc}
				\hline
				Method   &  \makecell{Storage \\ (MB/frame)}$\downarrow$ & \makecell{Training Time\\(s/frame)}$\downarrow$ & FPS$\uparrow$ \\
				\hline
				4DGS~\cite{yang2023gs4d}\qquad\qquad & 34.79 & \colorbox{best1}{50} & 56\\ 
                Spacetime Gaussian~\cite{li2023spacetime}     & 11.46 & 130 & \colorbox{best3}{69} \\
                DualGS~\cite{jiang2024robust} &  \colorbox{best1}{0.411} & 600 & \colorbox{best2}{83}\\
				HiFi4G~\cite{jiang2023hifi4g}  & 1.848 & 178 & 55\\
				$V^3$~\cite{wang2024v}  & \colorbox{best2}{0.753} & \colorbox{best2}{56} & 64\\
    		\hline
                Ours(Before Compression)  & 52.76 & \colorbox{best3}{108} & \colorbox{best1}{125}\\
                Ours(After Compression) & \colorbox{best3}{1.335} & \colorbox{best3}{108} & 65 \\
				\hline
			\end{tabular}
		}
		\label{table:comparison2}
    	\vspace{-10pt}
	\end{center}
\end{table}

\subsection{Ablations} \label{sec:abla} 

\paragraph{Candidate Filtering.} We conduct a qualitative ablation to evaluate our candidate filtering strategy in stable and dynamic topologies. As shown in Fig.~\ref{eval:tracker}, we visualize newly added Gaussians (in red) between the start and end frames.
Without the temporal tracker, redundant Gaussians are added in high-motion areas due to correspondence errors. Without photometric filtering, occluded regions are unnecessarily densified. Our full model suppresses redundancy and densifies only in areas with new observations.

\gcc{
\paragraph{Motion Gaussians Initialization.}
We compare point cloud initialization strategies for motion Gaussians, including random uniform sampling, SfM~\cite{7780814}, and EDGS~\cite{kotovenko2025edgs}. As shown in Tab.~\ref{tab:motion_init}, our EDGS-based scheme achieves the highest quality while producing the same number of Gaussians.
}

\paragraph{Appearance Gaussians Initialization.}
We compare initialization strategies for appearance Gaussians, including edge-based initialization with different neighbor counts at $4\times$, $9\times$, $16\times$, and random sampling at $9\times$ following the 3DGS~\cite{kerbl20233d} splitting strategy. As shown in Tab.~\ref{tab:appearance_init}, our $9\times$ edge-based scheme achieves a good trade-off between Gaussian count and quality, achieving high visual quality with reasonable points, comparable to the $16\times$ variant.

\begin{table}[t]
\centering
\caption{\gcc{Quantitative comparison of different initialization strategies for motion Gaussians.}}
\label{tab:motion_init}
\vspace{-9pt}
\resizebox{\columnwidth}{!}{
\begin{tabular}{l|c|c|c|c}
\hline
Configuration & PSNR $\uparrow$ & SSIM $\uparrow$ & LPIPS $\downarrow$ & Points Number $\downarrow$ \\
\hline
Uniform Sampling & 36.865 & 0.9883 & 0.0169 & 125342 \\
SfM & 36.902 & 0.9887 & 0.0172 & 127773 \\
\rowcolor{gray!50}
EDGS & 37.082 & 0.9900 & 0.0156 & 121048 \\
\hline
\end{tabular}
}
\vspace{-5pt}
\end{table}

\begin{table}[t]
\centering
\caption{Quantitative comparison of different initialization strategies for appearance Gaussians.}
\label{tab:appearance_init}
\vspace{-9pt}
\resizebox{\columnwidth}{!}{
\begin{tabular}{l|c|c|c|c}
\hline
Configuration & PSNR $\uparrow$ & SSIM $\uparrow$ & LPIPS $\downarrow$ & Points Number $\downarrow$ \\
\hline
4x (edge-based) & 36.879 & 0.9889 & 0.0180 & 53799 \\
\rowcolor{gray!50}
9x (edge-based) & 37.082 & 0.9900 & 0.0156 & 121048 \\
16x (edge-based) & 37.158 & 0.9903 & 0.0153 & 215197 \\
9x (random) & 36.929 & 0.9891 & 0.0154 & 121048 \\
\hline
\end{tabular}
}
\vspace{-9pt}
\end{table}

\begin{table}[t]
\centering
\caption{Per-frame quality and storage under different Quantization Parameter (QP) values for three sorting strategies.}
\label{tab:qp_storage}
\vspace{-9pt}
\resizebox{\columnwidth}{!}{
\begin{tabular}{c|cc|cc|cc}
\hline
\multirow{2}{*}{QP} & \multicolumn{2}{c|}{Ours (combined sorting)} & \multicolumn{2}{c|}{Lifespan-only sorting} & \multicolumn{2}{c}{Morton-only sorting} \\
\cline{2-7}
 & PSNR $\uparrow$ & Storage (KB) $\downarrow$ & PSNR $\uparrow$ & Storage (KB) $\downarrow$ & PSNR $\uparrow$ & Storage (KB) $\downarrow$ \\
\hline
5  & 36.507 & 2529.28 & 36.504 & 2595.23 & 36.512 & 3040.87 \\
15 & \cellcolor{gray!30}36.059 & \cellcolor{gray!30}1351.68 & 36.024 & 1413.12 & 36.068 & 1653.76 \\
25 & 31.837 & 783.36  & 31.817 & 826.88  & 31.888 & 1003.52 \\
\hline
\end{tabular}
}
\vspace{-11pt}
\end{table}

\paragraph{Sorting stategy.}
We conduct an ablation study to evaluate the impact of sorting strategies on the compressibility of 2D Gaussian attribute streams using standard video codecs. As shown in Tab.~\ref{tab:qp_storage}, we compare three configurations: our combined sorting strategy, Lifespan-only sorting, and Morton-only sorting. Our method consistently yields the lowest storage at the same quantization levels, benefiting from alignment with both intra-frame and inter-frame prediction mechanisms.

\subsection{Limitations and Discussions} 
While \abbtitle performs well in modeling dynamic scenes with topological changes, several limitations remain.
First, deactivated motion Gaussians cannot be reactivated, which may lead to redundancy. Leveraging video semantics could improve reuse and motion continuity.
\
Second, Gaussian removal is based solely on opacity, which may leave residual artifacts in occluded or unobserved regions. Semantic-aware deletion could enhance compactness.
\gcc{
In addition, \abbtitle relies heavily on CoTracker tracking and 2D segmentation masks. Tracking errors in textureless regions can introduce noticeable artifacts, while inaccurate masks may miss fine structures such as thin strings or hair. Incorporating 4D scene understanding could help address these issues.
Finally, as our approach is designed for indoor studio multi-view settings, it is not equipped to handle complex outdoor scenes with lighting variations, occlusions, and noise. Integrating inverse rendering techniques could help mitigate these challenges.}

\section{CONCLUSION}

We have presented \abbtitle, a novel dynamic Gaussian representation for robust tracking and high-fidelity rendering of human-centric 4D scenes with topological changes. By introducing a topology-aware motion-to-appearance framework, guided by a spatio-temporal tracker and photometric cues, our method adaptively detects new observations, while maintaining training efficiency through sparse motion anchoring and appearance Gaussians activation. The resulting representation supports long-term scene modeling with minimal overhead and aligns naturally with standard video codec pipelines for efficient compression.
Experiments demonstrate that \abbtitle achieves high visual quality across diverse dynamic scenes, while offering up to 40× compression. This makes it a practical and scalable solution for immersive volumetric video applications. We believe \abbtitle offers a solid step toward general-purpose, topology-aware 4D digitization of the physical world.

\bibliographystyle{ACM-Reference-Format}

\end{document}